

\hsize=14.cm
\vsize=21.cm   
\hoffset=1.2cm
\voffset=1.cm

\baselineskip=13pt plus 1pt minus 1pt  
\tolerance=1000
\parskip=0pt
\parindent=15pt

\def\normal{\baselineskip=13pt plus 1pt minus 1pt}

\nopagenumbers

\font\tenss=cmss10

\font\bfBig=cmb10  scaled\magstep2

\font\eightrm=cmr8
\font\ninerm=cmr9

\font\nineit=cmti9
\font\eightit=cmti8
\font\sevenmit=cmmi7

\font\eightsl=cmsl8

\font\ninebf=cmbx9

\def\myref#1{$^{#1}$}


\def\hspace{~\hskip 1cm ~}

\def\big{\displaystyle \strut }

\def\N{\kappa}

\def\vp{ {\vec p} }

\def\p{ {\bf p} }

\def\ZZ{ \hbox{\tenss Z} \kern-.4em \hbox{\tenss Z} }

\def\L{ {\cal L} }
\def\M{ {\cal M} }
\def\E{ {\cal E} }

\def\ep{\epsilon}
\def\eps{\varepsilon^{\mu\nu\rho}}
\def\d{\partial}

\def\tr{ {\,{\rm tr}\,} }
\def\Tr{ {\,{\rm Tr}\,} }

\def\d{\partial}
\def\la{\raise.16ex\hbox{$\langle$} \, }
\def\ra{\, \raise.16ex\hbox{$\rangle$} }
\def\st{\, \raise.16ex\hbox{$|$} \, }
\def\go{\rightarrow}

\def\tot{ {\rm tot} }

\def\psibar{ \psi \kern-.65em\raise.6em\hbox{$-$} }
\def\Lbar{ {\cal L} \kern-.65em\raise.6em\hbox{$-$} }


\global\newcount\refno
\global\refno=1 \newwrite\reffile
\newwrite\refmac
\newlinechar=`\^^J
\def \ref#1#2{\the\refno\nref#1{#2}}
\def\nref#1#2{\xdef#1{\the\refno}%
 \ifnum\refno=1\immediate\openout\reffile=refs.tmp\fi%
 \immediate\write\reffile{\noexpand\item{\noexpand#1.}#2\noexpand\nobreak}
 \immediate\write\refmac{\def\noexpand#1{\the\refno}}%
 \global\advance\refno by1}


\def\semi{ ^^J}                   

\def\refn#1#2{\nref#1{#2}}


\def\ijmpB#1#2#3{{\nineit Int.\ J.\ Mod.\ Phys.} {\ninebf {B#1}}, #3 (19{#2})}

\def\mplB#1#2#3{{\nineit Mod.\ Phys.\ Lett.} {\ninebf B{#1}}, #3 (19{#2})}
\def\plB#1#2#3{{\nineit Phys.\ Lett.} {\ninebf {#1}B}, #3 (19{#2})}

\def\np#1#2#3{{\nineit Nucl.\ Phys.} {\ninebf B{#1}}, #3 (19{#2})}
\def\prl#1#2#3{{\nineit Phys.\ Rev.\ Lett.} {\ninebf #1}, #3 (19{#2})}
\def\prB#1#2#3{{\nineit Phys.\ Rev.} {\ninebf B{#1}}, #3 (19{#2})}
\def\prD#1#2#3{{\nineit Phys.\ Rev.} {\ninebf D{#1}}, #3 (19{#2})}

\def\ptp#1#2#3{{\nineit Prog.\ Theoret.\ Phys.} {\ninebf {#1}}, #3 (19{#2})}

\def\cline{\hfil\noexpand\break  ^^J}

\immediate\openout\refmac=refno.tex

\refn\Redrich{
A.N.\ Redrich, \prl {52} {84} {18}; \prD {29} {84} {2366};
\semi K. Ishikawa, \prl {53} {84} {1615}; \prD {31} {85} {1432}.}

\refn\ChiralOne{
T.\ Appelquist, D.\ Nash, and L.C.R.\ Wijewardhana, \prl {60} {88} {2575};
\semi D.\ Nash, \prl {62} {89} {3024};
\semi T.\ Appelquist and D.\ Nash, \prl{64} {90} {721};
\semi R.\ Pisarski, \prD {44} {91} {1866};
\semi M.R.\ Pennington and D.\ Walsh, \plB {253} {91} {313};
\semi D.C.\ Curtis, M.R.\ Pennington, and D.\ Walsh, \plB {295} {92} {313};
\semi M.C.\ Diamantini, P.\ Sodano, and G.W.\ Semenoff, \prl {70} {93} {3848}.
}

\refn\ChiralTwo{
E.\ Dagotto, J.\ Kogut, and A.\ Kocic, \prl {62} {88} {1083}.}

\refn\electron{
A.\ Zee, \ijmpB {5} {91} {529}; \mplB {5} {91} {1339};
\semi P. Wiegmann, {\nineit Topological superconductivity}, Chicago preprint;
\semi E. Fradkin, (private communication).}

\refn\Hall{
S.C.\ Zhang, T.\ Hansson, and S.\ Kivelson, \prl {62} {89} {82};
\semi J.\ Jain, \prl {63} {89} {199}; \prB {40} {89} {8079};
\semi A.\ Balatsky and E.\ Fradkin, \prB {43} {91} {10622};
\semi A.\ Lopez and E.\ Fradkin, \prB {44} {91} {5246}.}

\refn\Landau{
See e.g.\ V.B.\ Berestetskii, E.M.\ Lifshitz, and L.P.\ Pitaevskii,
{\nineit ``Quantum Electrodynamics''}  (Pergamon Press, 1982),
section 32.}

\refn\Hosotani{
Computations here are parallel to those in the anyon superconductivity.
We follow the notation in
 Y.\ Hosotani, \ijmpB {7} {93} {2219}.}

\refn\Abrikosov{
See e.g.\ A.A.\ Abrikosov, L.P.\ Gorkov, and I.E.\ Dzyaloshinski, {\nineit
``Methods of Quantum Field Theory in Statistical Physics''} (Dover
Publications, 1975).}

\refn\Daemi{
S.\ Randjbar-Daemi, A.\ Salam, and J.\ Strathdee, \np {340} {90} {403};
and ref. \Hosotani.}

\refn\Feynman{
Useful relations  are (8-10) and (8-12),  with typos there corrected,
in R.P.\ Feynman and
A.R.\ Hibbs, {\nineit ``Quantum Mechanics and Path Integrals''} (McGraw-Hill,
1965).}


\def\smallkappa{\hbox{\sevenmit\char'24}}

\refn\Pauli{
Non-invariant (chirally asymmetric) Pauli-Villars regularization
is unnatural in the \smallkappa$\go$0 limit.}

\refn\Coleman{
S.\ Coleman and B.\ Hill, \plB {159} {85} {184};
\semi H.\ So, \ptp {74} {85} {585};
\semi N.\ Imai, K.\ Ishikawa, T.\ Matsuyama, and I.\ Tanaka, \prB {42}
{90} {10610}.}


\global\newcount\secno \global\secno=0
\global\newcount\appno \global\appno=0
\global\newcount\meqno \global\meqno=1
\global\newcount\figno \global\figno=1
\newwrite\eqmac
\def\eqn#1{
        \ifnum\secno>0
            \eqno(\the\secno.\the\meqno)\xdef#1{\the\secno.\the\meqno}%
            \immediate\write\eqmac{\def\noexpand#1{\the\secno.\the\meqno}}%
        \else\ifnum\appno>0
                 \eqno(\romappno.\the\meqno)\xdef#1{\romappno.\the\meqno}%
                \immediate\write\eqmac{\def\noexpand#1{\romappno.\the\meqno}}%
               \else
                  \eqno(\the\meqno)\xdef#1{\the\meqno}%
            \immediate\write\eqmac{\def\noexpand#1{\the\meqno}}%
                    \fi
        \fi
        \global\advance\meqno by1
          }
\newwrite\figmac
\def\fig#1{\ifnum\secno>0
            \the\figno\xdef#1{\the\figno}%
        \fi
        \global\advance\figno by1
          }

\immediate\openout\eqmac=lorentz.eq

\def\firstheadline{\hfil}
\def\otherheadline{\nineit
  \ifodd\pageno \qquad \hss Spontaneously Broken Lorentz
        Invariance \hss Page \folio
    \else Page \folio\hss Yutaka Hosotani \hss\qquad\fi}
\headline={\ifnum\pageno=1 \firstheadline \else\otherheadline\fi}



\baselineskip=9pt
\line{\ninerm Preprint from \hfil   UMN-TH-1211/93}
\line{\ninerm University of Minnesota \hfil August 6, 1993}

\vskip 2.cm

\baselineskip=20pt

\centerline{\bfBig Spontaneously Broken Lorentz Invariance}
\centerline{\bfBig in Three-Dimensional Gauge Theories}

\vskip 1.5cm

\baselineskip=12pt

\centerline{\ninerm  YUTAKA HOSOTANI}
\centerline{\eightit School of Physics and Astronomy, University of Minnesota}
\centerline{\eightit Minneapolis, MN 55455, U.S.A.}

\vskip .4cm

\centerline{\eightsl Type-set by plain \TeX }

\vskip .6cm

\baselineskip=10pt

{\eightrm
\midinsert \narrower\narrower   
In a wide class of three-dimensional Abelian gauge theories  with a bare
Chern-Simons term, the Lorentz invariance is spontaneously broken by
dynamical generation of a non-vanishing magnetic field.  A detailed computation
of an  energy density of the true vacuum is given.  The originally massive
photon becomes massless, fulfilling
the role of a Nambu-Goldstone boson associated with the spontaneous breaking
of the Lorentz invariance.
 \endinsert
}


\vskip .6cm
\normal

\secno=0  \meqno=1

Gauge theory in (2+1)-dimensions is rich.
A decade ago it was shown that in a general QED model a Chern-Simons term
is generated by radiative corrections so that a photon becomes topologically
massive.\myref{\Redrich}  It also has been argued, by both analytical
 and numerical  methods,
that in a T-symmetric  QED chiral symmetry is dynamically
broken.\myref{\ChiralOne,\ChiralTwo}

In this paper we would like to show that in a wide class of Abelian gauge
theories with a bare Chern-Simons term it is energetically favored
to generate a magnetic field, thus the Lorentz invariance being
spontaneously broken.
The true vacuum is a Hall liquid of fermions.  A photon
becomes massless, fulfilling the role of a Nambu-Goldstone boson.  The model
is important in describing  electron systems on a two-dimensional lattice
near half filling.\myref{\electron}  The mechanism of  dynamical
generation of a magnetic field is important in non-relativistic context as
well,
particularly in the fractional quantum Hall effect.\myref{\Hall}

Consider
$$\eqalign{
\L = &- {1\over 4} \, F_{\mu\nu}F^{\mu\nu} - {\N\over 2} \,
\eps A_\mu \d_\nu A_\rho   \cr
\noalign{\kern 4pt}
&+ \sum_a  {1\over 2} \, \big[ \,\psibar_a \, , \,
  \big( \gamma^\mu_a (i \d_\mu + q_a A_\mu)
    - m_a \big) \psi_a \, \big] ~. \cr}  \eqn\modelOne $$
Here $\psi_a$ is a two-component Dirac spinor.  In (2+1)-dimensions there
are two kinds of spinors, corresponding to the signature of two-dimensional
Dirac matrices
$$\eta_a = {i\over 2} \, \Tr \gamma_a^0\gamma_a^1\gamma_a^2 = \pm 1~.
   \eqn\signature  $$
We shall take a representation $\gamma^\mu_a = (\eta_a \sigma_3, i \sigma_1,
i\sigma_2)$ where $\sigma_j$'s are Pauli matrices.  The model is
invariant under charge conjugation.    Further, the transformation $m_a \go -
m_a$ is  equivalent to the transformation $\gamma^\mu_a \go - \gamma^\mu_a$ or
$\eta_a \go -\eta_a$.  Hence one can take $q_a>0$ and $m_a>0$ without loss
of generality.

Suppose that a magnetic field $B$ is spontaneously generated. We split
$A_\mu= A_\mu^{(0)} + A_\mu^{(1)}$ where
$A^{\mu(0)} = - B x_2 \, \delta^{\mu 1}$.
The Dirac equation in the presence of a constant magnetic field is exactly
solved.\myref{\Landau}  Energy eigenvalues are
$$\eqalign{
E^2 =  \omega_n^2 ~~~,~~~
\omega_n &= \sqrt{ m^2 + {2n\over l^2} } ~~~(n=0,1,2, \cdots) \cr
}   \eqn\eigenvalue  $$
where $l$ is the magnetic length:  $1/l^2 =|q B|$.
We have suppressed the species index $a$.

There is asymmetry in the lowest Landau level ($n$=0).  Depending on the signs
of $\eta$ and $qB$, either positive or negative energy states  exsit.
For instance,  for $\eta=+$ and $q B >0$,  eigenfunctions are
$$\eqalign{
u_{0p} =& {1\over (lL_1)^{1/2}} \, e^{-i\omega_0 t -ikx_1} \,
  \left[ \matrix{ v_0(\xi) \cr 0 \cr} \right]  \cr
\noalign{\kern 4pt}
u_{np} =& {1\over (lL_1)^{1/2}} \, e^{-i\omega_n t -ikx_1} \,
  {1\over \sqrt{2\omega_n} }
  \left[ \matrix{ \sqrt{ \omega_n + m} ~ v_n(\xi) \cr
      -i\sqrt{\omega_n-m} ~ v_{n-1}(\xi)  \cr} \right] \quad (n \ge 1) \cr
\noalign{\kern 4pt}
w_{np} =& {1\over (lL_1)^{1/2}} \, e^{+i\omega_n t -ikx_1} \,
  {1\over \sqrt{2\omega_n} }
  \left[ \matrix{ \sqrt{ \omega_n - m} ~ v_n(\xi) \cr
      +i\sqrt{\omega_n+m} ~ v_{n-1}(\xi)  \cr} \right] \quad (n \ge 1) \cr}
    \eqn\eigenfunction  $$
Here $k$=$2\pi p/L_1$ and  $\xi$=$(x_2/ l)$$-$$kl$ where $p$ is an integer.
$v_n(z)$=$(-1)^n 2^{-n/2} \pi^{-1/4} (n!)^{-1/2}$
$e^{z^2/2}$$(d^n  e^{-z^2} / dz^n)$
is the $n$-th eigenfunction in the harmonic oscillator
problem.  We have adopted a periodic boundary condition in the
$x_1$-direction, understanding the $L_1 \go \infty$ limit at the end.
The spatial part of the wave function of the $n$=0 mode is independent of mass
$m$.\myref{\Hosotani}

Solutions for $qB$$<$0 are obtained by charge conjugation, namely by
$\{ u^c(x)= U_c {\bar u(x)}^t$, $w^c(x)\}$ where $U_c=\gamma^2$.  Solutions for
$\eta=-$ are obtained by the transformation $t \go -t$ from the corresponding
ones in the $\eta=+$ case.

In the expansion of $\psi(x)$ one associates annihilation (creation) operators
for positive (negative) energy solutions.   For instance, for
$\eta=+$,
$$
\psi(x) = \sum_{np}  a_{np}^{} \,
   \left\{ \matrix{ u_{np}(x) \cr w_{np}^c(x) \cr} \right\}
 + \sum_{np}  b_{np}^\dagger \,
   \left\{ \matrix{ w_{np}(x) \cr u_{np}^c(x) \cr} \right\}
   \qquad {\rm for} ~ \left\{ \matrix{ qB >0 \cr qB <0 \cr} \right.
   \eqn\expansion  $$
with normal anticommutation relations among
$\{a_{np}, b_{np} \}$.    The summation over $n$ runs
from 0 or 1 for the $u_{np}$ or $w_{np}$ term, respectively.

In the massless limit the perturbative ground state with
$B \not= 0$ is infinitely degenerate.  As $\omega_0$=0, it costs no energy to
put a fermion in the lowest Landau level.   The degeneracy
is lifted by radiative corrections.  In this paper we consider two typical
states,  for sufficiently small masses,
 with the lowest Landau level  in each speceis being
empty ($\nu=0$) or completely filled ($\nu=1$).
The case of fractional filling will be
examined in a separate paper.  It will be found that a state
of the mixture of $\nu=0$ and $\nu=1$ among various speceis has a lower energy
than the  perturbative vaccum with $B=0$.

The charge $Q= \int d^2x ~ {1\over 2} \,q \, [\psi^\dagger ,\psi]$ is found,
by inserting (\expansion), to be
$$Q  = q \sum_p  \left\{ \matrix{
  a_{0p}^\dagger a_{0p}^{} - {1\over 2} \cr
  - b_{0p}^\dagger b_{0p}^{} + {1\over 2} \cr} \right\}
 + q \sum_{n=1}^\infty \sum_p
  ( a_{np}^\dagger a_{np}^{} - b_{np}^\dagger b_{np}^{} )~~.
       \eqn\charge $$
Here the upper and lower components correspond to $\eta\, \ep (B)>0$ and
$<0$, respectively.

The time component of
$\d_\nu F^{\nu\mu} - \N\, \eps \d_\nu A_\rho = j^\mu$
and the relation (\charge) imply that
$$\N \, B = \la j^0 \ra =
\sum_a \eta_a \, \ep (B) \, q_a (\nu_a - \hbox{${1\over 2}$})
\cdot { q_a |B|\over 2\pi}  ~~.
    \eqn\constraintOne    $$
Hence for $B\not= 0$
$$ \N  = {1\over 2\pi} \sum_a  \eta_a q_a^2 \,
   (\nu_a - \hbox{${1\over 2}$})  ~~~.  \eqn\constraintTwo  $$
This necessary condition, as is shown below, is related to
the Nambu-Goldstone theorem.  The value of $B$ is to be determined
by minimizing the energy density.

We need to compare the energy density of the perturbative vacuum with $B=0$,
$\E_0(e)$, with that of the non-perturbative vacuum,
$\E(e; q_aB, \nu_a)$.  Here we have collectively characterised the interaction
coupling $A_\mu^{(1)} j^\mu$ by $e$, supposing that all $q_a$'s in the
fluctuation part are proportional to  $e$.   Fermions couple to a dynamically
generated magnetic field in the combination $q_a B$ so that the
energy  density in the nonperturbative vacuum may be regarded as
a function of $e$, $\{ q_aB \}$, and filling factors $\{ \nu_a \}$.

It is beneficial to compare these energy densities with the corresponding
ones when the fluctuation coupling $e$ is turned off, while
$\{ q_aB \}$ kept fixed.  We write
$$\eqalign{
\Delta \E =& \E(e; q_aB,\nu_a) - \E_0(e)  \cr
\noalign{\kern 5pt}
=&  \Big\{ \E(e;q_aB,\nu_a) - \E(0;q_aB,\nu_a) \Big\}
- \Big\{ \E_0(e) - \E_0(0) \Big\} \cr
&\hskip 3.6cm  + \Big\{ \E(0;q_aB,\nu_a) - \E_0(0) \Big\} \cr
\noalign{\kern 5pt}
\equiv& \Delta \E^{(1)} - \Delta \E^{(2)} + \Delta \E^{(3)}  \cr}
   \eqn\EdifferenceOne  $$
$\Delta \E^{(3)}$ is the difference in energy densities in  the ``free''
 ($e$=0) theories. It is easy to see
$$\Delta \E^{(3)} = {1\over 2} \, B^2
  + \sum_a  {1 \over 2^{5/2} \pi^2}  ~ \zeta( \hbox{${3\over 2}$} )\,
\big| q_a B \big|^{3/2}   \eqn\EdiffTwo  $$
in the massless fermion limit.
The first term is the Maxwell energy, whereas the second comes from
the shift in zero point energies of fermions.  $\Delta \E^{(3)}$ is always
minimized at $B=0$.

$\Delta \E^{(1)}$ and $\Delta \E^{(2)}$  represent the shift in energy
densities when the fluctuation interaction $A^{(1)}_\mu j^\mu$ is turned on.
We employ an exact formula\myref{\Abrikosov}
$$
\Delta \E^{(1)} = i \int_0^e {de\over e}
 \int {d^3p\over (2\pi)^3} ~  \tr D_0^{-1} (p)
  \big\{ D(p~; q_aB,\nu_a) - D_0(p) \big\}   ~~~.
 \eqn\EshiftOne $$
Here  $D_0^{\mu\nu}(p)$ and $D^{\mu\nu}(p~;q_aB,\nu_a)$ are
the bare and full photon propagators in the nonperturbative vacuum.
The trace is taken over Lorentz indices.  For
$\Delta \E^{(2)}$,  $D(p~; q_aB,\nu_a)$ is replaced by $D(p~; B=0)$.
$D_0(p)$, independent of $B$, is given by
$$D^{\mu\nu}_0(p) = {1\over p^2 - \N^2} ~
\Big\{ g^{\mu\nu} - \N^2{p^\mu p^\nu \over  (p^2)^2}
   - i\N \, \eps\, {p_\rho\over p^2} \Big\} ~~.  \eqn\freePropagator $$

$D$ is related to
one-particle irreducible diagrams, $\Gamma$, by $D^{-1}= D_0^{-1} - \Gamma$.
The  rotational invariance and current conservation imply that\myref{\Daemi}
$$\eqalign{
\Gamma^{00} &= {\vp \,}^2 \Pi_0 \cr
\Gamma^{0j} &= p_0 p_j \Pi_0 - i \ep^{jk} p_k \Pi_1 \cr
\Gamma^{j0} &= p_0 p_j \Pi_0 + i \ep^{jk} p_k \Pi_1 \cr
\Gamma^{jk} &= \delta^{jk} p_0^2 \Pi_0 + i \ep^{jk} p_0 \Pi_1
   - ({\vp \,}^2 \delta^{jk} - p_j p_k) \Pi_2  \cr}
   \eqn\decomposition  $$
All $\Pi_j$'s are functions of $|p_0|$ and $|\vp\, |$.  (Here $p_j \equiv
p^j$)

If $\Gamma(p)$ is approximated by O($e^2$) diagrams, then the integration over
$e$ in (\EshiftOne) can be trivially performed, yielding
$$\eqalign{
\Delta \E^{(1)} - \Delta \E^{(2)} &=
-{i \over 2}  \int {d^3p\over (2\pi)^3} ~  \Big\{ \,
 \tr  \ln \,( 1 - \Gamma^{(2)} D_0) \big|_{qB,\nu} - (B \go 0) \, \Big\} \cr
\noalign{\kern 6pt}
&=-{i \over 2}  \int {d^3p\over (2\pi)^3} ~
\ln { (1+\Pi_0) \bigg\{ 1 + \displaystyle {1\over p^2} (p_0^2 \Pi_0
   - {\vp\,}^2 \Pi_2 ) \bigg\}    - {1\over p^2} (\N - \Pi_1)^2 \over
    (B\go 0 ) }   ~.   \cr}
 \eqn\EshiftTwo $$
If $\Pi_j(p)$ is further approximated by $\Pi_j(0)$, then
the above formula reduces to the one-loop effective action.  However,
we shall see below that it is absolutely necessary and crucial to take
full account of the momentum dependence.

To find $\Pi_j$, we need a fermion propagator,
$S(x,y)=-i \la T[\psi(x) \psibar(y)] \ra $, in the nonperturbative
vacuum.  First we note that
$$\eqalign{
S(x,y)_{B>0} &=e^{-i(x_1 - y_1) (x_2+y_2) /2l^2} \cdot S_0(x-y) ~~~, \cr
S_0(p)_{\eta=-} &= S_0(-p_0, \vp\,)_{\eta=+} ~~~, \cr
S(x,y)_{B<0} &= U_c S(y,x)^t_{B>0} U_c^\dagger ~~~, \cr}
   \eqn\Sone   $$
which follows from (\eigenfunction) and (\expansion).   It is straightforward
to see\myref{\Feynman} that
$$\eqalign{
S_0(p)^{\eta=+}_{\nu=0} &= -il^2 \int_0^\infty dt\, (\cos t)^{-1}  \,
   e^{i(p_0^2 - m^2 + i\ep) l^2 t - i {\vp\,}^2 l^2 \tan t}  \cr
&\hskip 1.cm \times \left[ \matrix{
   (m+p_0) e^{it} & -i(p_1 - ip_2) (\cos t)^{-1} \cr
   -i(p_1 + ip_2) (\cos t)^{-1} & (m-p_0) e^{-it} \cr} \right]  \cr
\noalign{\kern 6pt}
S_0(p)^{\eta=+}_{\nu=1} &= S_0(p)^{\eta=+}_{\nu=0}
  + 4\pi i \, e^{-{\vp\,}^2 l^2} \, \delta(p_0 - m)
   \left[ \matrix{ 1&0\cr 0&0\cr} \right]  ~. \cr}
    \eqn\Stwo  $$
 In the $B \go 0$ ($l \go \infty$) limit, $S_0(p)$ reduces to
$(p \gamma -m+ i\ep)^{-1}$.

$\Gamma^{(2)\mu\nu}(p)$ is given by
$$\Gamma^{(2)\mu\nu}(p) = iq^2 \int{d^3k\over (2\pi)^3} \,
\tr \gamma^\mu S_0(k) \gamma^\nu S_0(k-p) ~~~. \eqn\GammaOne  $$
To O($e^2$) a contribution from each fermion adds up to the total
$\Pi_j$.   An immediate consequence of (\decomposition), (\Sone), and
(\GammaOne) is, for each fermion loop,
$$\eqalign{
\Pi_j^{\eta=-} &= \pm \Pi_j^{\eta=+} \quad
 {\rm for~~} \Big\{ \matrix{j=0,2\cr j=1\cr}  \cr
\noalign{\kern 3pt}
\Pi_j^{B<0} &=  \Pi_j^{B>0} ~~~.  \cr}  \eqn\PiIdentity $$

The $\Pi_0$ and $\Pi_2$ parts of
the integral (\GammaOne) contain linear divergences, which need to be
regularized.  We employ the invariant Pauli-Villars regularization,
i.e.\ for each fermion loop with either $\eta$=+ or $-$, we associate
two regulator fields of $\eta$=$+$ and $-$, with equal weight, and let
the regulator masses go to $\infty$ at the end.\myref{\Pauli}
Divergent parts are independent of a mass $m$ and magnetic field $B$.
Straightforward manipulations lead to
$$\eqalign{
&\left[ \matrix{ \Pi_0 \cr \Pi_1 \cr p_0^2 \Pi_0 - {\vp\,}^2 \Pi_2 \cr}
  \right]_{\eta=+ , \nu=0} =
{q^2 l e^{i\pi/4} \over 8 \pi^{3/2} } \int_0^1 dw \int_0^\infty ds
{}~ \left[ \matrix{ f_0 \cr f_1 \cr f_2 \cr}  \right]  \cr
\noalign{\kern 6pt}
&\hskip 1cm \times \exp \bigg( {i\over 4} (1-w^2) s p_0^2 l^2
 - {i\over 2} {\cos ws - \cos s \over \sin s} \, {\vp\,}^2 l^2
 -im^2 l^2 s - \ep s \bigg)  \cr} \eqn\PiFormulaOne $$
where
$$\eqalign{
f_0 &= {s^{1/2}\over \sin s} \Big( 1+ {\cos s\over \sin s} {d\over ds} \Big)
(\cos ws - \cos s) \cr
\noalign{\kern 5pt}
f_1 &= - 2m {s^{1/2} \over \sin s} \, \cos ws  \cr
\noalign{\kern 5pt}
f_2 &= - s^{1/2} {d\over ds}\Big( {\sin ws\over \sin s} \Big)
   \cdot (wp_0^2 - {\sin ws \over \sin s} \, {\vp\,}^2 )
   +{s^{1/2} \cos ws \over \sin s} (1-w^2) p_0^2 \cr
&\hskip 2.5cm - {s^{1/2}\over \sin s} \Big( {1\over \sin s}
+ \cos ws \,{d\over ds} \Big) \Big({\cos ws - \cos s \over\sin s} \Big)
  \cdot {\vp\,}^2 ~~.  \cr}   \eqn\PiFormulaTwo  $$
For $\nu=1$
$$\eqalign{
&\left[\matrix{ \Pi_0 \cr \Pi_1 \cr \Pi_2 \cr} \right]_{{\eta=+ \atop \nu=1}}
=\left[\matrix{ \Pi_0 \cr \Pi_1 \cr \Pi_2 \cr} \right]_{{\eta=+ \atop \nu=0}}
     + {q^2\over \pi} \int_0^\infty ds ~
\left[ \matrix{ \displaystyle -(p_0)^{-1} \sin 2mp_0 l^2 s \cr
     i \cos 2mp_0 l^2 s \cr 0 \cr}              \right]   \cr
&\hskip 3.cm  \times \exp \Big( -2is + ip_0^2 l^2 s
  - {{\vp\,}^2 l^2\over 2}  (1- e^{-2is}) - \ep s \Big) ~.  \cr}
   \eqn\PiFormulaThree $$

The induced Chern-Simons coefficient,
$- \Pi_1$ at $p^\mu=0$, can be evaluated exactly.
$$\eqalign{
\Pi_1(0) \big|_{B=0} &= - \eta \, {q^2\over 4\pi} \cr
\Pi_1(0) \big|_{B\not=0} &=  \eta \, (\nu- {1\over 2})  {q^2\over 2\pi} \cr}
   \eqn\PiOneValueOne $$
The sign of $\Pi_1(0)$ flips when the lowest Landau level is
filled ($\nu=1$).   We also note that
$\lim_{m\go 0} \Pi_1(p^2$$\not=$$0)\big|_{B=0}$=0
so that two limits $m\go 0$ and $p\go 0$ do not commute with each other.
The necessary condition (\constraintTwo) is expressed as
$$\Pi_1(0) = \N ~~~, \eqn\constraintThree $$
i.e.\ the bare Chern-Simons term must be exactly cancelled by the induced
Chern-Simons term in order to have $B\not=0$.

These functions are  not analytic in $B$.
The integrals in the $B\go 0$ ($l \go \infty$) limit are
dominated at small $s$.  Hence behavior at small $|B|$ is found by expanding
$\cos s$ in the exponent and others in Taylor series in $s$.    At $B=0$
(\PiFormulaOne) and (\PiFormulaTwo) reproduce the standard result:
$$\eqalign{
\Pi_0 \big|_{B=0}  &=  \Pi_2 \big|_{B=0}
={q^2\over 8\pi} {1\over (-p^2)^{1/2} } \,
   \Big( \sqrt{z}  + (1-z)   \sin^{-1} {1\over \sqrt{1+z}} \Big)  \cr
\noalign{\kern 5pt}
\Pi_1\big|_{B=0} &= - {q^2\over 4\pi}  \sqrt{z} \,
  \sin^{-1} {1\over \sqrt{1+z}} \cr}
   \eqn\PiFormulaBzero  $$
where $z=4m^2/(-p^2)$.

Recalling $q|B|=1/l^2$, we see that
$$\Pi_j(p)\big|_{B,\nu=0} = \Pi_j(p)\big|_{B=0} + {\rm O}(B^2)  ~~.
   \eqn\smallBone $$
When the lowest level is filled,
$$\eqalign{
\Pi_0(p)\big|_{B,\nu=1} - \Pi_0(p)\big|_{B=0}
&\sim {q^2\over 2\pi} {1\over p_0} \, \Big\{ {1\over 2-p^2 l^2 - 2mp_0 l^2}
   - {1\over 2-p^2 l^2 + 2mp_0 l^2} \Big\}  \cr
\Pi_1(p)\big|_{B,\nu=1} - \Pi_1(p)\big|_{B=0}
&\sim \eta \, {q^2\over 2\pi}  \, \Big\{ {1\over 2-p^2 l^2 - 2mp_0 l^2}
   + {1\over 2-p^2 l^2 + 2mp_0 l^2} \Big\}  ~~~. \cr}
       \eqn\smallBtwo  $$
$\Pi_2|_{\nu=1} = \Pi_2|_{\nu=0}$ as noted above.   For small $m$
a significant correction of O($|B|$) arises only for $\Pi_1(p)$ at $\nu=1$.

The leading term in (\EshiftTwo), and therefore in $\Delta \E$ in
(\EdifferenceOne),  is of order $|B|$.  In a wide class of models
 the sign of the coefficient turns out to be
negative so that the energy density is lowered by $B\not=0$.

Since the integrand in (\EshiftTwo) behaves like $\ln p^2$ (or $1/p^4$)
for small (or large) $p$, the integral converges.
In particular, one can Wick-rotate the $p_0$ integral.  Hence in terms of
$\p$=$(\vp ,p_3=-ip_0)$
$$\Delta \E^{(1)} - \Delta \E^{(2)} =
{1 \over 2}  \int {d^3\p \over (2\pi)^3} ~
\ln { (1+\Pi_0) \bigg\{ 1 + \displaystyle {1\over \p^2} (p_3^2 \Pi_0
   + {\vp\,}^2 \Pi_2 ) \bigg\}  + {1\over \p^2} (\N - \Pi_1)^2 \over
    (B\go 0 ) }   ~~~.
 \eqn\EshiftThree $$

It is in order to give a simple model.  Suppose that a system is chirally
symmetric, i.e.\ there are an equal number  of
$\eta$=+ and $-$ fermions ($N_f^+$=$N_f^-$=$N_f$),
each pair having the same mass and charge.
Because of (\PiIdentity), $\Pi_1^\tot(p) \big|_{B=0} =0$.
If all fermions have the same charge $q_a$=$e$, and
all $\eta$=+ ($-$) fermions have $\nu$=1 ($\nu$=0), the condition
(\constraintTwo) or (\constraintThree) reads
$$ \N = {e^2\over 2\pi} \, N_f ~~. \eqn\constraintFour $$
In the massless fermion limit
$$\Pi_1^\tot(p)\Big|_{B\not=0} \sim
 {e^2N_f\over 2\pi} \, {2\over 2+ \p^2 l^2} ~~. \eqn\smallBthree  $$
$\Pi_0^\tot$ and $\Pi_2^\tot$ are given by $2N_f$ times the formula in
(\PiFormulaBzero) up to O($B^2$) corrections.

Hence in the argument of the logarithm in (\EshiftThree), the numerator is
always smaller than the denominator, implying that the energy difference
is negative.  One finds
$$\Delta \E
= -    {N_f e^2\over 2\pi^3} \tan^{-1}{4\over \pi}\cdot e|B|
 + {\rm O}(|B|^{3/2}) ~~~.  \eqn\EshiftFour  $$
The energy density is minimized at $B\not= 0$,  the Lorentz invariance being
spontaneously broken.  The true vacuum is a Hall liquid with $\nu$=1 for
$\eta$=1 fermions.   If $\N<0$, the condition is satisfied by having
$\nu$=1 ($\nu$=0) for $\eta$=$-$ (+).
Although the sign of $B$ is not determined in this approximation,
higher order corrections will lift the degeneracy   if $\N\not= 0$.

As a generalization one can consider a case where $\N$=$(e^2/2\pi) N_0$
($N_0$: a positive integer), $N_f^+$=$N_f^-$=$N_f> N_0$.  If $N_0$ fermions of
$\eta$=+ have $\nu$=1 and all others have $\nu$=0,
the two conditions, (\constraintTwo) and $\Delta\E(B\not=0)<0$, are
satisfied.

In the case of QED, namely with $\N=0$, the situation is opposite.
If the fermion content is chirally symmetric so that
$\Pi_1^\tot(p)|_{B=0}$=0, then $\nu$=1 fermions give positive contributions
of O($|B|$) to $\Delta\E$.  The energy is minimized at $B$=0, which is
consistent with the assertion that chiral symmetry is dynamically
broken.\myref{\ChiralOne,\ChiralTwo}
If the fermion content is not chirally symmetric, the sign of an O($|B|$) term
in $\Delta\E$ depends on both $\{ \nu_a \}$ and $\{ m_a \}$, and
a detailed numerical evaluation is necessary to determine the sign.

Finally we would like to clarify the relation between the condition
(\constraintThree) and Nambu-Goldstone theorem.   We first note that the
location of poles of the photon propagator is determined by zeros of
$$\det D^{-1}(p) = (p^2)^2  \bigg\{ (1+\Pi_0)
( p^2 + p_0^2 \Pi_0 - {\vp\,}^2 \Pi_2) - (\N- \Pi_1)^2 \bigg\} ~~.
      \eqn\poleOne$$
The front factor $(p^2)^2$ represents gauge degrees of freedom, while
the factor in the braket determines the photon spectrum $p_0(\vp\,)$.
The condition (\constraintThree) implies that
$$\lim_{\vp \go 0} p_0(\vp\,) = 0 ~~, \eqn\poleTwo $$
i.e.\ a photon is a massless excitation.

On the other hand we have an identity
$$
\lim_{p\go 0} \int d^3x \, e^{ipx} \d_\rho^x
\la {\rm T} [\M^{0j\rho}(x) F_{0k}(0) ] \ra
 =  i \ep^{jk} \la F_{12} (0) \ra  ~~.   \eqn\NGtheorem  $$
Here $\M^{\mu\nu\rho}=x^\mu T^{\nu\rho} - x^\nu T^{\mu \rho}$ where
$T^{\mu\nu}$ is the symmetric energy-momentum tensor.
$M^{0j} = \int d^2x\,\M^{0j0}$ generates a Lorentz boost.
(\NGtheorem) implies that a dynamically generated $B$=$-\la F_{12}\ra$$\not=$0
must accompany a massless excitation which couples to $\M^{0j\rho}$ and
$F_{0k}$.   It is nothing but a photon.  Its massless nature,
(\poleTwo), is guaranteed by the Nambu-Goldstone theorem associated with the
spontaneous breaking of the Lorentz invariance.  Consequently the relation
(\constraintThree) should acquire no higher order corrections.\myref{\Coleman}

Under a Lorentz boost the new vacuum is transformed into a  state with
$B$$\not=$0 and $F_{0j}$$\not=$0.  The transformed state has a higher energy,
being unstable against pair creation.   There is no degeneracy associated with
the spontaneous symmetry breaking.  The Nambu-Goldstone boson is a vector,
but not a scalar.

In this paper we have shown that the Lorentz invariance is spontaneously
broken in a wide class of models by dynamical generation of a magnetic
field.   We have considered only integer filling ($\nu$=0 or 1),
in which case the
mechanism works only for a quantized bare Chern-Simons coefficient.
For a general value of the coefficient, one needs to consider fermion
states of fractional filling.  This, with applications to non-relativistic
systems, will be discussed separately.

\vfil\eject

\baselineskip=12pt
\centerline{\ninebf Acknowledgements}

{\ninerm
\midinsert\narrower
This work was supported in part by the U.S.\ Department of Energy under
contract no.\ DE-AC02-83ER-40105.  The author would like to thank Valery
Rubakov, Yannick Meurice, Koichi Yamawaki, and Eduardo Fradkin for many
enlightening discussions at the early or final stage of investigation.  He is
grateful to the Aspen Center for Physics for its hospitality, where a part of
work was done. \endinsert
}

\baselineskip=12pt

\centerline{\ninebf References}
\bigskip

\ninerm
 \immediate\closeout\reffile
	\input refs.tmp\vfill\eject\nonfrenchspacing

\bye